\documentclass[
superscriptaddress,
aps
]{revtex4-2}

\usepackage{graphicx}
\usepackage{amsmath,amssymb}
\usepackage{paralist, tabularx}
\usepackage{dcolumn}
\usepackage{bm}
\usepackage{textgreek}
\usepackage{xcolor,hyperref}
\definecolor{darkblue}{rgb}{0.0,0.0,0.5}
\definecolor{darkgrey}{rgb}{0.5,0.1,0.1}
\hypersetup{colorlinks,breaklinks,
            linkcolor=darkblue,urlcolor=darkblue,
            anchorcolor=darkblue,citecolor=darkblue}

\usepackage{xspace}
\newcommand{\nav}{Na\textsubscript{V}1.1\xspace}  
\newcommand{\na}{Na\textsuperscript{+}\xspace}  
\usepackage{siunitx}
\newenvironment{nalign}{
    \begin{equation}
    \begin{aligned}
}{
    \end{aligned}
    \end{equation}
    \ignorespacesafterend
}

\begin{document}

\title{Depolarization block induction via slow \texorpdfstring{\nav}{NaV1.1} inactivation in Dravet syndrome}
\author{Louisiane Lemaire}
\email{louisiane.lemaire@inria.fr}
\affiliation{MathNeuro team, Inria Branch at the University of Montpellier, Montpellier, 34090, France
}
\author{Mathieu Desroches}
\affiliation{MathNeuro team, Inria Branch at the University of Montpellier, Montpellier, 34090, France
}
\author{Serafim Rodrigues}
\affiliation{MCEN research group, BCAM - Basque Center for Applied Mathematics, Bilbao, 48009, Basque Country, Spain
}
\affiliation{Ikerbasque - The Basque Foundation for Science, Bilbao, 48009, Basque Country, Spain
}
\author{Fabien Campillo}
\email{fabien.campillo@inria.fr}
\affiliation{MathNeuro team, Inria Branch at the University of Montpellier, Montpellier, 34090, France
}

\begin{abstract}
Dravet syndrome is a developmental and epileptic encephalopathy, characterized by the early onset of drug-resistant seizures and various comorbidities. Most cases of this severe and complex pathology are due to mutations of \nav, a sodium channel mainly expressed in fast-spiking inhibitory neurons. Layer et al. (\textit{Front. Cell. Neurosci.} \textbf{15}, 2021) showed that one of these mutations alters the voltage dependence of channel activation, as well as the voltage dependence and kinetics of slow inactivation. Implementing the three effects into a computational model, they predict that altered activation has the largest impact on channel function, as it causes the most severe firing rate reduction. Using a conductance-based model tailored to the dynamics of fast-spiking inhibitory neurons, we look deeper into slow inactivation. We exploit the timescale difference between this very slow process and the rest of the system to conduct a multiple-timescale analysis. We find that, upon prolonged stimulation, the onset of slow inactivation at lower voltage in mutant channels promotes depolarization block, another possible firing deficit aside from frequency reduction. The accelerated kinetics of slow inactivation in mutant channels hastens this transition. This suggests that slow inactivation alterations might for some Dravet variant contribute to the pathological mechanism.
\end{abstract}

\maketitle

\section*{\textbf{Introduction}}
Dravet syndrome \cite{dravet1978epilepsies} (DS) is a severe and complex form of epilepsy. It is classified as a developmental and epileptic encephalopathy (DEE), i.e. a condition characterized by the early onset of epileptic seizures on a background of developmental impairment that tends to worsen as a consequence of epilepsy \cite{guerriniDevelopmentalEpilepticEncephalopathies2023,schefferILAEClassificationEpilepsies2017}.
Cognitive, motor and behavioral development are affected, as well as sleep quality \cite{mantegazzaSodiumChannelopathiesSkeletal2021}. 
In particular, DS is frequently associated with autistic-like behavior \cite{liAutismDravetSyndrome2011}.
The first seizure typically occurs during the first year of life in a child with no pathological history, often in the context of hyperthermia (i.e., a febrile seizure)
\cite{dravetChapter65Dravet2013}. This marks the start of the febrile stage, first of the three stages described by C. Dravet \cite{dravetChapter65Dravet2013}, during which seizures -- whether febrile or not -- occur at a moderate frequency. First-line anticonvulsants often fail to stop them \cite{dravetChapter65Dravet2013,gataullinaGenotypePhenotypeDravet2017}.
As the disease progresses into the worsening stage, seizure frequency intensifies, additional seizure types appear and developmental delay becomes more evident \cite{gataullinaGenotypePhenotypeDravet2017,alonsogomezNewTherapeuticStrategies2022}.
Between the ages of five and ten, the child enters the stabilization stage: seizures are less frequent and some seizure types may disappear \cite{gataullinaGenotypePhenotypeDravet2017, dravetChapter65Dravet2013}. However, both seizures and comorbidities persist in adulthood, requiring lifelong care \cite{gentonDravetSyndromeLongterm2011}.
Despite partially effective drugs such as stiripentol \cite{chironStiripentolSevereMyoclonic2000}, fenfluramine \cite{lagaeFenfluramineHydrochlorideTreatment2019,nabboutFenfluramineTreatmentResistantSeizures2020} or cannabidiol \cite{devinskyTrialCannabidiolDrugResistant2017} and promising novel approaches like disease-modifying and genetic therapies \cite{hanAntisenseOligonucleotidesIncrease2020,michInterneuronspecificDualAAVSCN1A2025,colasanteDCas9BasedScn1aGene2020,tanenhausCellSelectiveAdenoAssociatedVirusMediated2022},
DS remains drug resistant for most patients \cite{rusinaVoltagegatedSodiumChannels2023,wirrellInternationalConsensusDiagnosis2022}.
The prevalence of early mortality is estimated at approximately 10 to 15\% \cite{sakauchiRetrospectiveMultiinstitutionalStudy2011,cooperMortalityDravetSyndrome2016}, primarily caused by sudden unexpected death in epilepsy (SUDEP) or status epilepticus \cite{sakauchiRetrospectiveMultiinstitutionalStudy2011,shmuelyMortalityDravetSyndrome2016}.

DS has a genetic cause \cite{bergRevisedTerminologyConcepts2010}. Most cases (at least $80\%$) are due to mutations of the \textit{SCN1A} gene \cite{claesNovoMutationsSodiumChannel2001,depienneSpectrumSCN1AGene2009,catterallDravetSyndromeSodium2018,harkinSpectrumSCN1ArelatedInfantile2007,zuberiGenotypePhenotypeAssociations2011}, which encodes the voltage-gated sodium channel \nav. 
With rare exceptions \cite{vanNovoHomozygousVariant2021}, all mutations are found in the heterozygous state \cite{pariharSCN1AGeneVariants2013} (i.e., DS patients still possess a non-mutated copy of \textit{SCN1A}) and most of them arise \textit{de novo} \cite{claesNovoMutationsSodiumChannel2001,harkinSpectrumSCN1ArelatedInfantile2007,mariniIdiopathicEpilepsiesSeizures2007} (i.e., they are absent in the parents). About 50-60\% of DS mutations are truncating mutations \cite{depienneSpectrumSCN1AGene2009,zuberiGenotypePhenotypeAssociations2011,ishiiClinicalImplications1A2017}, expected to give rise to a nonfunctional protein product \cite{rusinaVoltagegatedSodiumChannels2023}. They are thought to cause pure haploinsufficiency, i.e. a halving of the number of functional channels without negative dominance \cite{bechiPureHaploinsufficiencyDravet2012}. On the other hand, missense mutations (i.e., resulting in the substitution of an amino acid) may affect channel properties without necessarily leading to a complete abolition of its function \cite{layerDravetVariantSCN1AA1783V2021}. They represent about 40 to 50\% of DS mutations \cite{depienneSpectrumSCN1AGene2009,pariharSCN1AGeneVariants2013,zuberiGenotypePhenotypeAssociations2011,ishiiClinicalImplications1A2017}.

\nav channels are mainly expressed in GABAergic neurons, notably in parvalbumin-positive (PV+) fast-spiking ones, and are crucial to their excitability \cite{rusinaVoltagegatedSodiumChannels2023,gataullinaGenotypePhenotypeDravet2017,ogiwaraNav11LocalizesAxons2007,yuReducedSodiumCurrent2006,wangDevelopmentalChangesNav112011,duttonPreferentialInactivationScn1a2013a}. 
Hypoexcitability of these inhibitory neurons, due to \nav loss of function, has been proposed as the major pathological mechanism of DS, making circuits hyperexcitable and more prone to seizures \cite{berardinoGeneTherapyDravet2024}. Although disinhibition is still thought to play an important role during the initial phase of the disease \cite{rusinaVoltagegatedSodiumChannels2023}, the overall scenario is more complex. This initial defect is suspected to induce and be accompanied by secondary changes, both compensatory mechanisms such as the upregulation of \na channels (i.e., homeostatic remodeling) and the establishment of additional dysfunctions (i.e., pro-pathologic remodeling) \cite{rusinaVoltagegatedSodiumChannels2023}. In particular, it has been proposed that hyperexcitability of some glutamatergic neurons might participate to the DS phenotype at later stages of the disease \cite{mattisCorticohippocampalCircuitDysfunction2022,berardinoGeneTherapyDravet2024}. The implication of excitatory neurons is supported by studies using patient-derived induced pluripotent stem cells (iPSCs) \cite{liuDravetSyndromePatientderived2013,jiaoModelingDravetSyndrome2013}.

In this work, we explore in depth how the activity of fast-spiking GABAergic neurons is affected by the presence of mutant \nav channels. More precisely, we are interested in variants that alter the properties of \nav without suppressing its function as a channel. This case has been less studied than that of truncating mutations \cite{kuoDisorderedBreathingMouse2019}, for which mouse models have been available for nearly two decades \cite{yuReducedSodiumCurrent2006,ogiwaraNav11LocalizesAxons2007}.
In terms of disease progression, our analysis focuses on the time window following the replacement of Na\textsubscript{V}1.3 channels (embryonic Na\textsubscript{V} subtype) with \nav and preceding major remodeling effects. A precise characterization of the firing deficits of GABAergic neurons at this stage is important to investigate their potential contribution to secondary responses such as the pathological remodeling of excitatory networks.

We focus on the recurrent missense DS mutation A1783V, localized at the end of segment 6 in domain 4 of \nav \cite{mariniIdiopathicEpilepsiesSeizures2007}, for which knock-in mouse models are available \cite{ricobarazaEpilepsyNeuropsychiatricComorbidities2019,sattaNeuropathologicalCharacterizationDravet2021}. Based on recordings in a heterologous expression system, Layer \emph{et al.} \cite{layerDravetVariantSCN1AA1783V2021} showed that A1783V alters the voltage dependence of channel activation, as well as the voltage dependence and kinetics of slow inactivation. Slow inactivation is a mechanism distinct from the fast inactivation of sodium channels at each spike, developing much more slowly, during prolonged trains of depolarization \cite{vilinSlowInactivationVoltagegated2001}. Implementing the three effects of the mutation in a conductance-based model, Layer \emph{et al.} predicted that altered activation is the main mechanism underlying GABAergic neuron dysfunction, as it causes the most severe reduction in firing rate. 
They note an accelerated rundown of sodium current amplitudes due to altered slow inactivation, though this has only a marginal effect on firing frequency in the model. However, simulating over longer durations may be necessary to observe an impact of a process as slow as slow inactivation.
Using a conductance-based model tailored to the dynamics of fast-spiking inhibitory neurons in the dentate gyrus \cite{huComplementaryTuningNa2018}, we look deeper into slow inactivation, exploiting the timescale difference with the rest of the system.

Our main finding is that enhanced slow inactivation of mutant \na channels leads to an increased propensity of the GABAergic neuron to enter a depolarization block (DB) upon sustained electric current stimulation.
DB refers to a stable stationary state of the neuronal membrane in which it remains depolarized but electrically silent and unresponsive to further stimulation, a behavior attributed to the saturation or inactivation of \na channels. This neuronal state emerges when input currents exceed the spiking regime. For an illustration, see already Fig.~\ref{fig:bd_wrt_iapp} panel \textbf{a}, where the DB regime corresponds to the family of stable steady states that exist past the second Hopf bifurcation (see also \cite{izhikevichDynamicalSystemsNeuroscience2007a}). 
DB can be easily obtained in patch-clamp electrophysiology even in physiological conditions, upon sufficiently strong stimulation.
In the model, we find that implementing a shift upregulation of the voltage dependence of slow inactivation reduces the stimulation intensity required to trigger this transition. Moreover, implementing the accelerated kinetics of slow inactivation speeds up the transition.
In addition, the sensitivity to depolarization block is accentuated at elevated temperature, which is consistent with the occurrence of febrile seizures in Dravet patients. This inability to sustain tonic firing represents another potential firing deficit of GABAergic neurons, aside from frequency reduction. 

The rest of the article is organized as follows. We first present the model and dissect its fast and slow dynamics. Then, our results are organized in three sets of computational experiments, all aimed at investigating the propensity of the neuron to reach DB in case of prolonged activity. Specifically, we examine: i) the effect of implementing an enhanced slow inactivation for all \na channels, ii) the effect of implementing it only for half of the \na channels,  and iii) the effect of hyperthermia. Finally, we discuss our results and outline several directions for future work.
\section*{The model}
\subsection*{A conductance-based model of a fast-spiking GABAergic neuron}

Our work builds on the model proposed by Hu \emph{et al.}~\cite{huComplementaryTuningNa2018}. It is based on the Wang-Buzs{\'a}ki model~\cite{wangGammaOscillationSynaptic1996}, with its gating dynamics fitted to voltage-clamp data of fast-spiking PV+ GABAergic neurons of the dentate gyrus. These neurons are known to express \nav \cite{ogiwaraNav11LocalizesAxons2007,wangDevelopmentalChangesNav112011,duttonPreferentialInactivationScn1a2013a}, and the dentate gyrus is considered a key locus of DS pathology and seizure generation \cite{liautardHippocampalHyperexcitabilitySpecific2013, steinHippocampalDeletionNaV112019}, motivating our choice. Note that Hu \emph{et al.} recorded directly in the axons, where \nav expression was demonstrated \cite{ogiwaraNav11LocalizesAxons2007}.

Following Layer \emph{et al.} \cite{layerDravetVariantSCN1AA1783V2021}, we implemented in this model the slow inactivation of sodium channels, represented by the gating variable $s$.
To compensate for the reduction in sodium conductance compared to the model without slow inactivation, we increased $g_{\rm Na}$ by dividing it by the steady state level of slow inactivation $s_{\infty}$ at $\qty{-70}{\milli\volt}$, which, after rounding, gives $g_{Na} = \qty{70}{\milli\siemens\per\centi\meter\squared}$. 
The dependence of the gating rates on temperature is already accounted for in the model by Hu \emph{et al.} for $h$, $n$ and $\tilde{n}$. We model it in the same way for $s$.

The model's equations are given by:
\begin{equation}
	\label{eq:hu}
	\begin{split}
		C\,\dot{v}
		& = - I_{\rm Na} - I_{\rm K} - I_{\rm leak} + I_{\rm app}\,, 
		\\
		\dot{x}
		& = \Phi_x({\rm Temp})\,
		\bigl(x_{\infty}(v) - x\bigr)/\tau_x(v) \,,\;x\in\{h,n,\tilde{n}\}\,,\\
		\dot{s}
		& =   
		\Phi_s({\rm Temp})\,\bigl(s_{\infty}(v - v_s) -s \bigr)/\tau_{s} \,,
	\end{split}
\end{equation}
with the ionic currents:
\begin{equation}
	\begin{split}
	  I_{\rm Na}   
	  & = g_{\rm Na} \, m_{\infty}^3(v) \, h \, s\,(v-E_{\rm Na})\,,      
	  \\
	  I_{\rm K}    
	  & = 
	  g_{\rm K}\, n^3 \,\tilde{n} \,(v-E_{\rm K})\,,
	  \\
	  I_{\rm leak}    
	  & = g_{\rm leak} \, (v-E_{\rm leak})\,.
	\end{split}
\end{equation}
Note that Hu \textit{et al.} introduced the gating variable $\tilde{n}$, which, in combination with $n$, they argue offers a more accurate description of potassium channel activation.
The temperature dependence is modeled via the function: $\Phi_y({\rm Temp})=(Q_{10, y})^{\frac{{\rm Temp}-{\rm Temp}_y}{10}}$ for $y\in\{h,n,\tilde{n},s\}$ with ${\rm Temp}_h={\rm Temp}_n={\rm Temp}_{\tilde{n}}=24\,$\unit{\degreeCelsius} and ${\rm Temp}_s=33\,$\unit{\degreeCelsius}. The temperature values ${\rm Temp}_y$ correspond to the condition in which recordings were performed respectively by Hu \emph{et al.} and Layer \emph{et al.}
The voltage-dependent gating functions are given in the Supplementary Information (SI), as well as the chosen parameter values (see Table~S1).

A central aspect of this model is the inclusion of alterations of \nav gating dynamics, for which we also followed Layer \emph{et al.} \cite{layerDravetVariantSCN1AA1783V2021}. 
Based on recordings of tsA-201 cells expressing the \nav\textsuperscript{A1783V} variant, they observed three effects: a $\qty{10}{\milli\volt}$ depolarizing shift in the voltage dependence of activation, a $\qty{15}{\milli\volt}$ hyperpolarizing shift in the voltage dependence of slow inactivation, and a ten-fold acceleration of slow inactivation.
In this paper, we focus on the alterations of slow inactivation. We implemented them as in Layer \emph{et al.}, with a shift $v_s=\qty{-15}{\milli\volt}$ of $s_{\infty}$ and a tenfold smaller value of the time constant $\tau_s$.
Note that both alterations enhance slow inactivation, which corresponds to a loss of function of the channel.
At this point, we apply these changes to all sodium channels; in a subsequent section, we will examine what occurs when only a fraction of sodium channels is affected.

\subsection*{The model's slow-fast structure}
As is typical of Hodgkin-Huxley type models, System~\eqref{eq:hu} features several timescales, with for instance the membrane potential evolving faster than channel dynamics. There are also timescale differences among gating variables, particularly pronounced in our case due to the implementation of slow inactivation, on top of the classical fast inactivation of sodium channels. To gain a bird's-eye view of the different timescales at play in System~\eqref{eq:hu}, we first simulate the model and compare the time profile of its variables; see Fig.\,S1. Overall, the slow inactivation gating variable $s$ is much slower than the other state variables, reflecting the very large value of its time constant $\tau_s=\qty{30000}{\milli\second}$.
This behavior is still preserved when $\tau_s$ is decreased to the pathological value of $\qty{3000}{\milli\second}$ (Fig.~S2).
We exploit this timescale separation in the present work by considering four fast variables ($v$, $h$, $n$ and $\tilde{n}$) and one slow variable ($s$). 

The simplest way to exploit timescale separation in order to analyse a system with both fast and slow variables, is to consider the so-called \textit{fast subsystem}. It consists of keeping the fast equations intact while freezing the slow dynamics and considering slow variables (in our case $s$) as parameters that force the fast equations. 
Then, the bifurcation structure of the fast subsystem with respect to the slow variables provides key elements about the full system's dynamics. Namely, superimposing a full system's trajectory onto the fast subsystem's bifurcation diagram reveals that the trajectory slowly follows branches of attractors of the bifurcation diagram up to bifurcation points, near which it switches on the fast timescale to another branch of attractors. 
This procedure,  called \textit{slow-fast dissection}~\cite{rinzel87},  is a standard tool in analysing multiple-timescale systems; see already Fig.~\ref{fig:slow_fast_all_channels} panel {\bf c} for an illustration.

Parts of the full system trajectories that slowly follow a stable stationary branch of the fast subsystem's bifurcation diagram correspond to \textit{quiescent phases}. Parts that oscillate fast while the upper and lower envelopes of the oscillations slowly follow a stable periodic branch of the fast subsystem's bifurcation diagram are referred to as \textit{bursting phases}. 
Note that, during bursting phases, the dynamics of our slow variable $s$ is perturbed by the fast voltage spikes via the voltage dependence of $s_{\infty}$ (see insets of Figs. S1 and S2).
One can study such phases by introducing the \textit{averaged slow subsystem}~\cite{baer1995,roberts2017geometric}. In that limit, for values of $s$ such that there exists a stable limit cycle of the fast subsystem, the dynamic of the slow variable $s$ is averaged over one period $T(s)$ of that limit cycle:
\begin{equation}
	 \dot{\langle s\rangle} 
	 = 
	 \frac{\Phi_s({\rm Temp})}{\tau_s} \,
	 \left( \frac{1}{T(s)}  \int_0^{T(s)} s_{\infty} \bigl(v(s, \tau)-v_s\bigr) \; {\rm d}\tau - s \right).
\end{equation}
Intuitively, the averaged slow subsystem aims to analyze, in the singular limit, the slow drift that the system performs during the burst phase, which is obtained by averaging out the fast rotations. This limiting system is defined only on the manifold of limit cycles of the fast subsystem. This system allows us to determine whether, on average, $s$ decreases or increases and drives the membrane potential towards either tonic spiking or DB.
During quiescent phases, averaging is not necessary and the slow dynamics is given by the standard slow subsystem:
\begin{equation}
    \begin{split}
	 0&= - I_{\rm Na} - I_{\rm K} - I_{\rm leak} + I_{\rm app}\,,\\
	 0&=x_{\infty}(v)-x,\;x\in\{h,n,\tilde{n}\}\,,\\
	 \dot{s}&=
		\Phi_s({\rm Temp})\,\bigl(s_{\infty}(v - v_s) -s \bigr)/\tau_{s}
	\end{split}
\end{equation}
Next, we employ the mathematical and computational framework of multiple-timescale analysis to study the role of sodium channels' slow inactivation in mediating DB dynamics in System~\eqref{eq:hu}.
\section*{Results}

\subsection*{Enhancing slow inactivation for all \na channels}

To gain insight into how the neuron's activity is impacted by the two alterations of slow inactivation, we first assume they affect all \na channels. This is, of course, not realistic: even if we ignore other voltage-gated \na channels besides \nav, non-mutant \nav channels are synthesized via the expression of the wild type \textit{SCN1A} allele in heterozygous. 
However, this assumption allows us to understand the mechanisms in a simpler setting.

\begin{figure}[t!]
\centering
\includegraphics[]{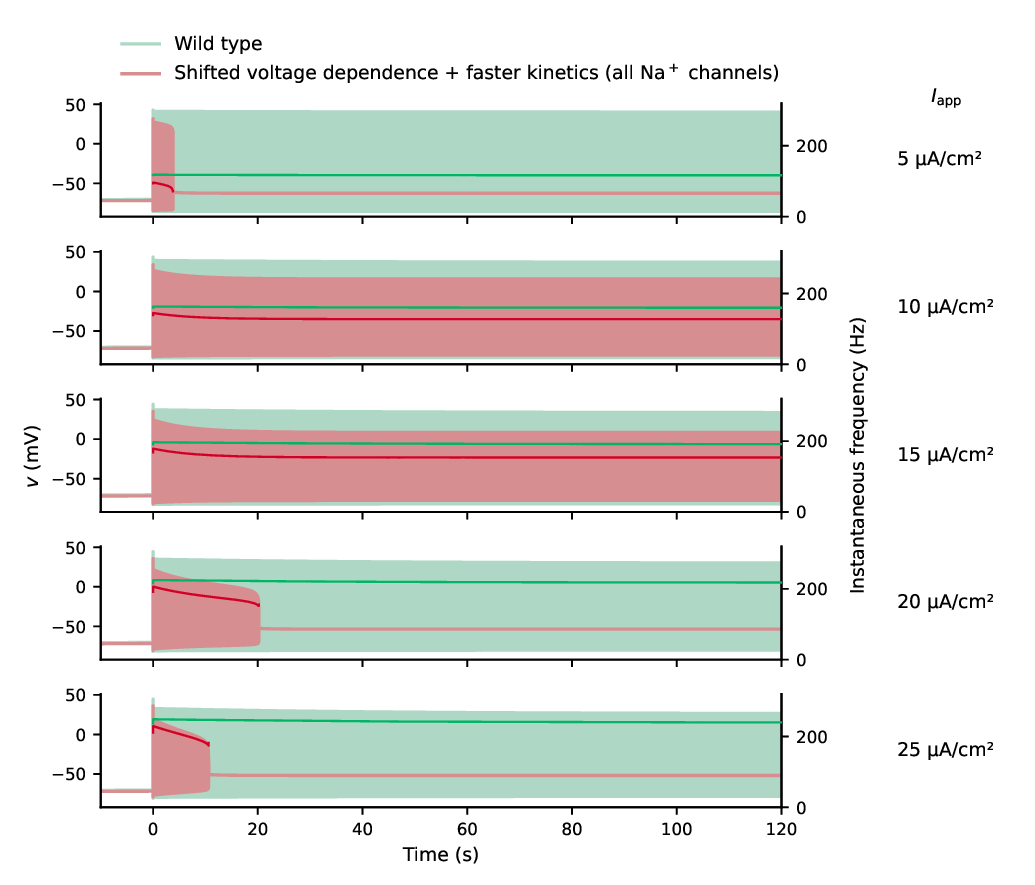}
\caption{Effect of the altered slow inactivation of all \na channels on the firing activity. We show voltage traces in response to two-minute currents steps of different intensities, starting the simulations from the resting state of the neuron (steady state when $I_{\rm app} = \qty{0}{\micro\ampere\per\centi\meter\squared}$). Condition 1 (green curves): wild type neuron, i.e. default gating dynamics for all \na channels. Condition 2 (red curves): enhanced slow inactivation
for all \na channels, i.e. shifted voltage dependence (${v}_s = \qty{-15}{\milli\volt}$) and faster kinetics ($\tau_{s, \rm mut} = \qty{3000}{\milli\second}$).}
\label{fig:voltage_traces_all_channels}
\end{figure}

Figure~\ref{fig:voltage_traces_all_channels} compares the cases without ($v_s=\qty{0}{\milli\volt}$, $\tau_s=\qty{30000}{\milli\second}$; green curves) or with ($v_s=\qty{-15}{\milli\volt}$, $\tau_s=\qty{3000}{\milli\second}$; red curves) the alterations of slow inactivation. Given that slow sodium inactivation takes time to develop, we perform simulations over two minutes. Voltage time traces are superimposed in response to five increasing values of the applied current $I_{\rm app}$. One can observe two main effects. 
The first and more radical effect of enhancing slow inactivation is the occurrence of a DB at large values of the applied current. Note that, at low current values (e.g., for $I_{\rm app} = \qty{5}{\micro\ampere\per\centi\meter\squared}$), when slow inactivation is enhanced, the neuron also enters a stationary regime after transiently spiking.
The mechanism underlying the cessation of firing, analyzed in more detail in subsequent sections, is similar at both low and large current values (see already Fig. S3), and relies on the slow build-up of slow inactivation.
However, the stationary regimes may be interpreted as either resting state or DB, depending on their location relative to the family of limit cycles in the bifurcation diagram with respect to $I_{\rm app}$ (see already Fig.~\ref{fig:bd_wrt_iapp} (\textbf{b})).
The second effect is a reduction in firing frequency compared with the wild type case, which becomes more pronounced over the course of the simulation, as shown by the darker colored lines indicating the evolution of the instantaneous firing frequency.
This frequency decay is accompanied by smaller spike amplitudes, mainly due to a reduction of the peak voltage (as opposed to minimum voltage). 

In order to understand these effects we exploit the timescale separation between the slow variable $s$ and the other variables of the model, focusing on the case $I_{\rm app} = \qty{20}{\micro\ampere\per\centi\meter\squared}$ (next to last panel of Fig.~\ref{fig:voltage_traces_all_channels}).
In a first step, we investigate the role of the parameter $v_s$, which controls the voltage dependence of slow inactivation, while keeping the time constant $\tau_s$ at its default value.
Subsequently, we implement the reduction of $\tau_s$ as well, thereby combining both alterations.

\subsubsection*{Slow inactivation developing at lower voltages}

\begin{figure}[t!]
\centering
\includegraphics[]{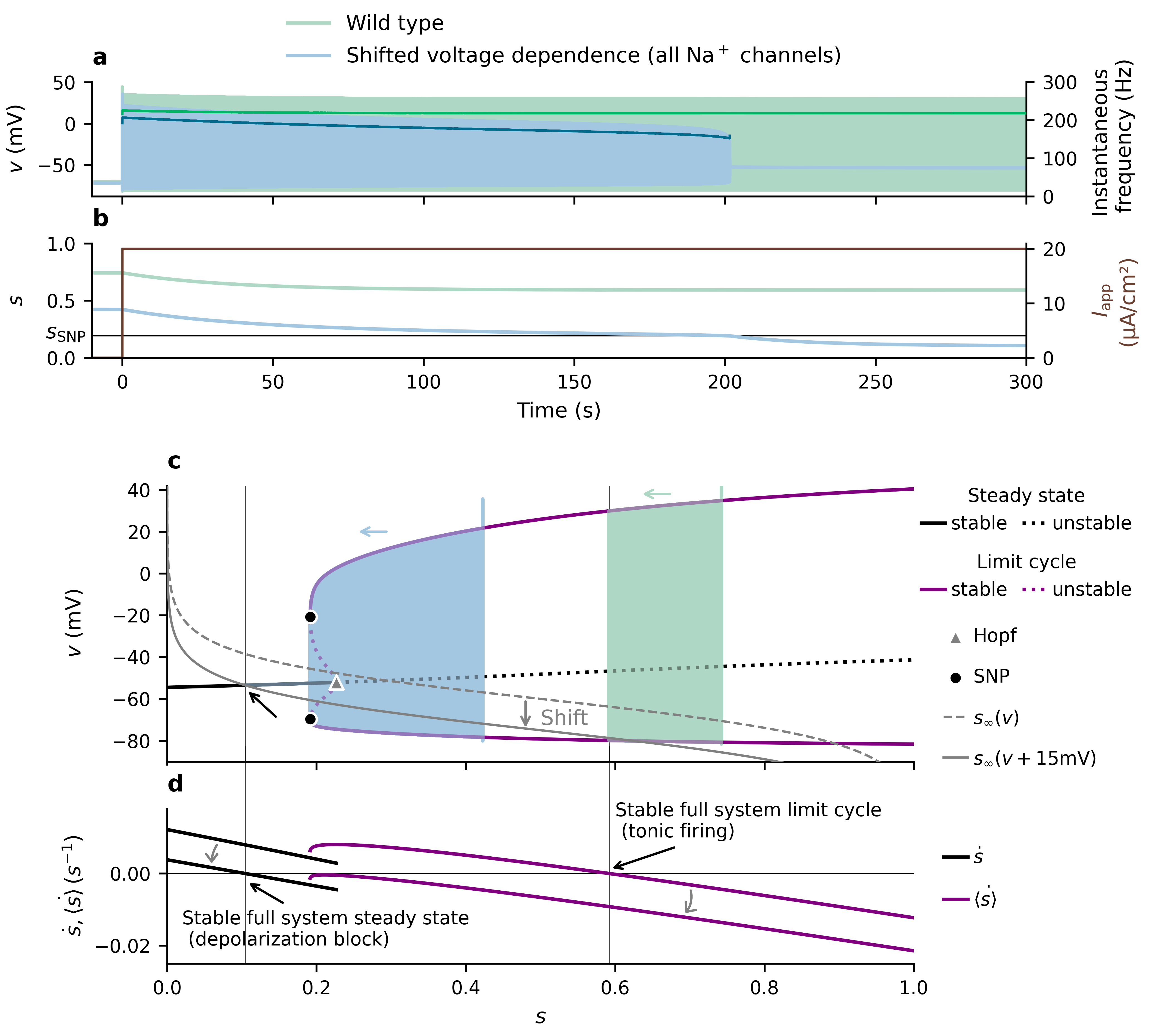}
\caption{The development of slow inactivation at lower voltage values can cause the neuron to eventually enter depolarization block; case study when $I_{\rm app} = \qty{20}{\micro\ampere\per\centi\meter\squared}$.
Voltage {\bf (a)} and $s$ {\bf (b)} time traces in response to a two-minute current step {\bf (b, right axis)}, in the wild type case (${v}_s = \qty{0}{\milli\volt}$; green curves) or with the shift of the voltage dependence of slow inactivation (${v}_s = \qty{-15}{\milli\volt}$; blue curves). {\bf (a, right axis)} Instantaneous firing frequency (darker colored lines).
{\bf (c)} Full system trajectories projected onto the bifurcation diagram of the fast subsystem with respect to $s$ when $I_{\rm app} = \qty{20}{\micro\ampere\per\centi\meter\squared}$.
{\bf (d)} Time derivative of the slow variable $s$ on stable steady states of the fast subsystem or averaged time derivative of $s$ on stable limit cycles of the fast subsystem, without (upper curves) or with (lower curves) the shift of $s_{\infty}$.
        }
\label{fig:slow_fast_all_channels}
\end{figure}

We start by shifting the voltage dependence of slow inactivation by $\qty{15}{\milli\volt}$.
In this scenario, in response to a step current of $\qty{20}{\micro\ampere\per\centi\meter\squared}$ (brown curve in panel \textbf{b} of Fig.~\ref{fig:slow_fast_all_channels}) the neuron eventually enters DB (blue curve in panel \textbf{a}), whereas the wild type neuron responds to the same stimulation with tonic firing (green curve in panel \textbf{a}).
In the first case $s$ converges to very low levels, indicating nearly complete inactivation of \na channels, while for the wild type neuron $s$ remains relatively high (blue vs. green curves in panel \textbf{b}).
Note that the initial values for $s$ already differ between the two conditions, since we start simulations from the resting states
and that, with the shift of $s_{\infty}$, the steady state value of $s$ is lower at the resting potential.

We now analyze the bifurcation diagram of the fast subsystem of System~\eqref{eq:hu} with respect to $s$ (now a parameter), shown in panel \textbf{b}.
This diagram is common to both conditions, since the only difference between them (the value of $v_s$) concerns the dynamics of $s$, here frozen.
Depending on the value of $s$, we find a stationary regime and a periodic regime separated by a subcritical Hopf bifurcation, giving rise to a family of initially unstable limit cycles, which restabilizes at a saddle-node of periodics (\textsf{SNP} on the figure). Then, applying Rinzel's slow-fast dissection, we superimpose onto this bifurcation diagram the trajectories of the full system whose voltage and $s$ time profiles are shown in panel \textbf{a}. 

In the wild type scenario (green trajectory), before the current step $s$ is approximately $0.75$. Note that when $I_{\rm app} = \qty{0}{\micro\ampere\per\centi\meter\squared}$, the bifurcation diagram of the fast subsystem is different from the one at $I_{\rm app} = \qty{20}{\micro\ampere\per\centi\meter\squared}$ (panel \textbf{c} of Fig.~S4 versus panel \textbf{c} of Fig.~\ref{fig:slow_fast_all_channels}).
When the step current is applied, the neuron starts to spike and \na channels overall slowly inactivate. Indeed, panel  \textbf{d} shows that in this region, $s$ decreases on average: $\dot{\langle s \rangle} < 0$.
The full system follows the family of stable limit cycles of the fast subsystem (panel \textbf{c}), until it reaches a stable equilibrium of the averaged slow subsystem (panel \textbf{d}), which corresponds to a stable limit cycle of full system. Hence, in this case the long-term behaviour of the system is tonic firing.

In contrast, when $s_\infty$ is shifted towards more hyperpolarized voltages, when $I_{\rm app} = \qty{0}{\micro\ampere\per\centi\meter\squared}$ the variable $s$ is already smaller than in the wild type case (see Fig.~S4). When the step current is applied, the neuron emits spikes while $s$ decreases on average. However, unlike in the wild type configuration, there is no equilibrium of the averaged slow subsystem (panel \textbf{d} of Fig.~\ref{fig:slow_fast_all_channels}).
Hence, as the neuron spikes, $s$ slowly decreases towards the SNP of the fast subsystem, where the trajectory enters into a quasi-stationary regime. At this stage, the dynamics is entirely slow. The trajectory approaches a stable steady state of the full system, identified as the intersection between the shifted function $s_\infty$ and the branch of stable steady states of the fast subsystem (which is clear from the equations).
This is confirmed by showing that $\dot{s}$ converges to $0$ at $s\approx0.11$ (see panel \textbf{d}). This stable equilibrium at high input current corresponds to a DB.

\subsubsection*{Faster slow inactivation}

\begin{figure}[ht!]
\centering
\includegraphics[width=0.99\linewidth]{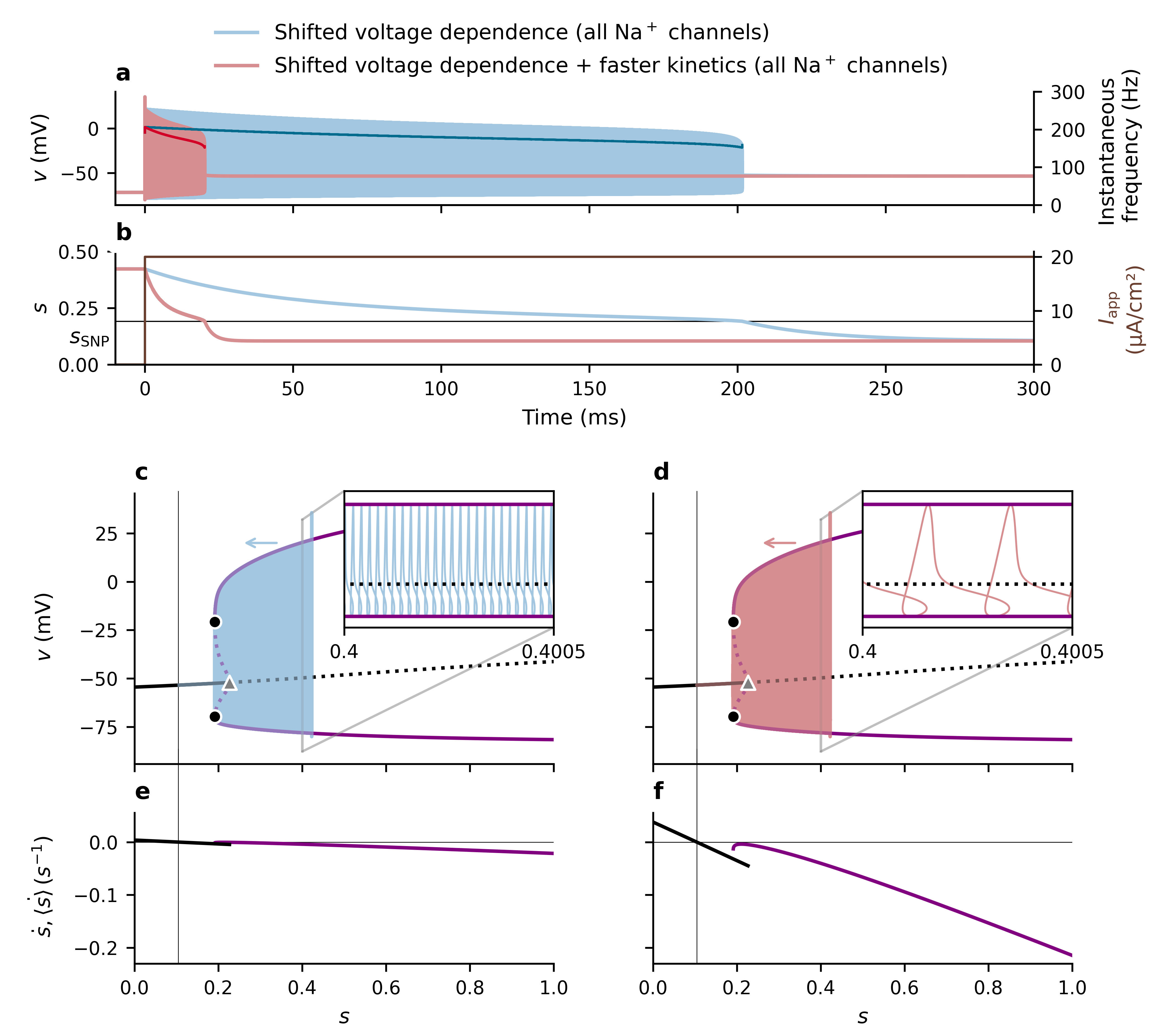}
\caption{
The accelerated kinetics of slow inactivation hastens the transition to DB; case study when $I_{\rm app} = \qty{20}{\micro\ampere\per\centi\meter\squared}$.
Voltage {\bf (a)} and $s$ {\bf (b)} time traces in response to a two-minute current step {\bf (b, right axis)}, with only the shift of slow inactivation voltage dependence and default kinetics (${v}_s = \qty{-15}{\milli\volt}$, $\tau_{s, \rm wt} = \qty{30000}{\milli\second}$; blue curves) or with both the shift of the voltage dependence and faster slow inactivation kinetics (${v}_s = \qty{-15}{\milli\volt}$, $\tau_{s, \rm wt} = \qty{3000}{\milli\second}$; red curves). {\bf (a, right axis)} Instantaneous firing frequency (darker colored lines).
{\bf (c-d)} Full system trajectories projected onto the bifurcation diagram of the fast subsystem with respect to $s$ when $I_{\rm app} = \qty{20}{\micro\ampere\per\centi\meter\squared}$, in each condition.
{\bf (e-f)} Time derivative of the slow variable $s$ on stable steady states of the fast subsystem or averaged time derivative of $s$ on stable limit cycles of the fast subsystem, in each condition. 
}
\label{fig:slow_fast_tau_all_channels}
\end{figure}

We now modify the kinetics of slow inactivation, in addition to the shift of the voltage dependence, by taking a smaller value for $\tau_{s}$. This additional change speeds up the kinetics of $s$, yet still slow compared to the other variables of the model. As a consequence, the slow-fast dissection is still valid, as shown in panels {\bf (c-d)} of Fig.~\ref{fig:slow_fast_tau_all_channels},
where one can clearly observe that the full system still follows stable branches of the fast subsystem. 
As $s$ decays (on average) faster, it drives the full system faster along the stable branch of limit cycles of the fast subsystem (panels {\bf (d, f)}), compared to the case where only the voltage dependence of slow inactivation is altered (panels {\bf (c, e)}).
As a result, the system reaches the SNP bifurcation sooner (panel {\bf (b)}). Hence, faster slow inactivation kinetics hastens the transition to DB (panel {\bf (a)}).

\subsection*{Enhancing slow inactivation for half of the \na channels}

We now apply the alterations of slow inactivation only to a fraction of the \na channels. Our goal is to assess the extent to which the effects observed in the previous section are maintained when a subset of the channels remain functionally normal. To implement this, following Layer \emph{et al.}, we replace the sodium current $I_{\rm Na}$ with the sum of a current $I_{\rm Na, wt}$ through wild type channels and a current $I_{\rm Na, mut}$ through altered channels, with:
\begin{equation}
	\begin{split}
	I_{\rm Na, wt}  
	  & = 
	  (1-p_{\rm mut}) \,g_{\rm Na} \,m_{\infty}^3 \,h \, s_{\rm wt} \,(v-E_{\rm Na})\,,
	  \\
	  I_{\rm Na, mut}  
	  & = 
	  p_{\rm mut} \,g_{\rm Na} \,m_{\infty}^3 \,h \, s_{\rm mut} \,(v-E_{\rm Na})\,,
	\end{split}
\end{equation}
where $p_{\rm mut}$ is the proportion of \na channels affected by the alterations. Here we take $p_{\rm mut}=0.5$. Hence, $s_{\rm wt}$ and $s_{\rm mut}$ are distinct slow inactivation gating variables, corresponding to the respective channel populations. Their dynamics is given by:
\begin{equation}
    \label{eq:two_s}
	\begin{split}
	\dot{s}_{\rm wt}
		& =   
		\Phi_s({\rm Temp})\,
		\bigl(s_{\infty}(v ) - s_{\rm wt}\bigr)/\tau_{s, {\rm wt}}\,,\\
	\dot{s}_{\rm mut}
		& =   
		\Phi_s({\rm Temp})\,
		\bigl(s_{\infty}(v - v_s) - s_{\rm mut}\bigr)/\tau_{s, {\rm mut}}\,,
	\end{split}
\end{equation}
with $\tau_{s, {\rm wt}}=\qty{30000}{\milli\second}$ and $\tau_{s, {\rm mut}}=\tau_{s, {\rm wt}}$ or $\tau_{s, {\rm mut}}=\qty{3000}{\milli\second}$.
Equations~\eqref{eq:two_s} replace the equation for $s$ in System~\eqref{eq:hu}. We therefore have a system with two slow variables instead of one, $s_{\rm mut}$ being in some configurations faster than $s_{\rm wt}$. The averaged slow subsystem is now two-dimensional:
\begin{equation}
    \label{eq:averaged_slow_two_s}
	\begin{split}
	\dot{\langle s_{\rm wt}\rangle}
		& =   
		\frac{\Phi_s({\rm Temp})}{\tau_{s, {\rm wt}}} \,
		\left( 
		   \int_0^{T(s_{\rm tot})}s_{\infty}\bigl(v(s_{\rm tot},\tau)\bigr) \; {\rm d}\tau - s_{\rm wt}
		\right),\\
	\dot{\langle s_{\rm mut}\rangle}
		& =   
		\frac{\Phi_s({\rm Temp})}{\tau_{s, {\rm mut}}} \,
		\left(\int_0^{T(s_{\rm tot})}s_{\infty}\bigl(v(s_{\rm tot},\tau) - v_s\bigr) \; {\rm d}\tau - s_{\rm mut}\right),
	\end{split}
\end{equation}
with $s_{\rm tot} = p_{\rm wt} \,s_{\rm wt} + p_{\rm mut} \, s_{\rm mut}$.

Note that the positive time constants $\tau_{s,{\rm wt}}$ and $\tau_{s,{\rm mut}}$ do not modify the steady states of System (\ref{eq:averaged_slow_two_s}) and their stability. The same is true for the steady states of the standard slow subsystem, which describes the dynamics of the slow variables on steady states of the fast subsystem. Therefore, as long as the timescale separation is strong enough, $\tau_{s, {\rm mut}}$ does not affect the long-term behavior of the full system, aside from the small fluctuations -- that average out at each spike -- of $s_{\rm wt}$ and $s_{\rm mut}$.
As when slow inactivation is altered for all \na channels (previous section), the time constant $\tau_{s, {\rm mut}}$ influences the transient dynamics, while the voltage shift $v_s$ can influence whether the neuron eventually enters DB.

\begin{figure}[ht!]
\centering
\includegraphics[]{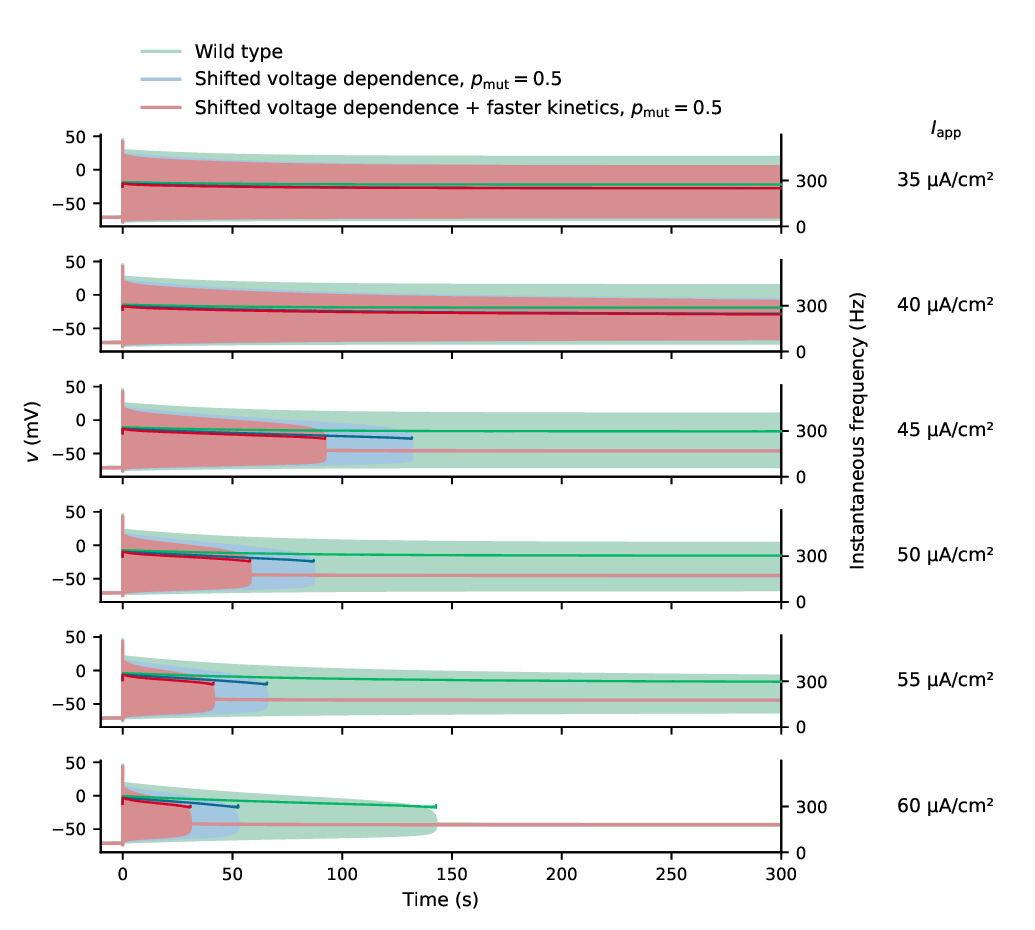}
\caption{
Effect of the altered slow inactivation of only half of the \na channels on the firing activity. We show voltage traces in response to five-minute currents steps of different intensities, starting the simulations from the resting state of the neuron (steady state when $I_{\rm app} = \qty{0}{\micro\ampere\per\centi\meter\squared}$). Condition 1 (green curves): wild type neuron, i.e. default gating dynamics for all \na channels ($p_{\rm mut}=0$). Condition 2 (blue curves): shifted voltage dependence of slow inactivation ($v_s = \qty{-15}{\milli\volt}$) for half of the \na channels ($p_{\rm mut}=0.5$); default kinetics ($\tau_{s, \rm mut} = \tau_{s, \rm wt} = \qty{30000}{\milli\second}$).  Condition 3 (red curves): shifted voltage dependence of slow inactivation ($v_s = \qty{-15}{\milli\volt}$) and faster kinetics ($\tau_{s, \rm mut} = \qty{3000}{\milli\second}$), for half of the \na channels ($p_{\rm mut}=0.5$).
}
\label{fig:voltage_traces_heterozygous}
\end{figure}

\begin{figure}[ht!]
\centering
\includegraphics[]{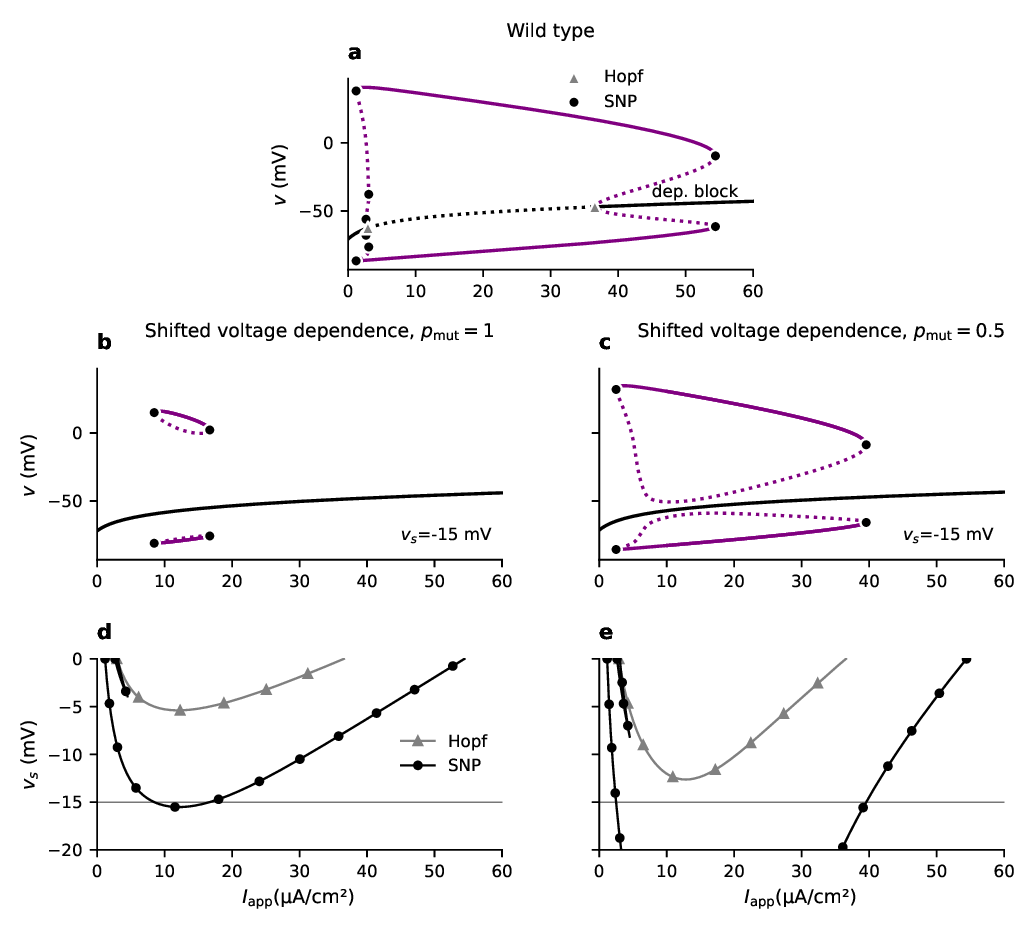}
\caption{Asymptotic behavior of the full system. {\bf (a-c)} Bifurcation diagrams of the full system with respect to the applied current. {\bf (a)} Wild type neuron ($p_{\rm mut}=0$). {\bf (b)} Shifted slow inactivation voltage dependence ($v_s = \qty{-15}{\milli\volt}$) for all \na channels ($p_{\rm mut}=1$). {\bf (c)} Shifted slow inactivation voltage dependence ($v_s = \qty{-15}{\milli\volt}$) for half of the \na channels ($p_{\rm mut}=0.5$). {\bf (d-e)} Two-parameter bifurcation diagrams with respect to the applied current and shift $v_s$, when $p_{\rm mut}=1$ (panel {\bf d}) and when $p_{\rm mut}=0.5$ (panel {\bf e}). 
}
\label{fig:bd_wrt_iapp}
\end{figure}

The voltage traces in Fig.~\ref{fig:voltage_traces_heterozygous} illustrate that shifting the voltage dependence of slow inactivation with $v_s$ lowers the applied current threshold for the transition to DB, and that faster slow inactivation kinetics precipitates this transition. 
These effects, which we explained using slow-fast dissection in the simpler case of the previous section (Figs.~\ref{fig:slow_fast_all_channels},\ref{fig:slow_fast_tau_all_channels}), are therefore preserved when only half of the sodium channels have altered gating. While applying slow-fast analysis in this more realistic configuration is feasible (see Fig. S5), it provides little additional insight. Instead, for a systematic overview of the long-term behavior of the neuron, we perform a numerical continuation of the full system with respect to the applied current (see Fig.~\ref{fig:bd_wrt_iapp} (\textbf{a-c})). When the shift $v_s$ becomes more negative, the two SNP bifurcations, which stabilize from either side the family of limit cycles of the full system, move closer together (\textbf{e}). This reduces the range of current intensities for which long-term tonic firing is possible, compared to the wild type case (panels \textbf{c} versus panel \textbf{a}).
Beyond the second SNP bifurcation (i.e., the one at higher applied current values), the neuron may spike transiently, even for tens of seconds, but it will ultimately enter DB.
As expected, the deformation of the bifurcation diagram with respect to $I_{\rm app}$ is not as pronounced as when the voltage dependence of slow inactivation is shifted for all \na channels (panels \textbf{b} and \textbf{d}, $p_{\rm mut} = 1$), but it is still present.
When $s_{\infty}$ is shifted for all \na channels, the two SNPs eventually merge at a point -- presumably an \emph{isola formation center} -- beyond which no limit cycle exists in the full system.
This happens at a value of $v_s$ close to that measured experimentally \cite{layerDravetVariantSCN1AA1783V2021}, placing it within a biologically realistic range.
On the other hand, when $s_{\infty}$ is shifted for only half of the \na channels, stable limit cycles of the full system seem to persist regardless of how negative $v_s$ becomes (not shown), due to the presence of normally functioning channels.

Note that as $v_s$ becomes more negative, the two Hopf bifurcations giving rise to the family of limit cycles also move closer to each other (grey lines in panels \textbf{d} and \textbf{e}). Eventually, they collide and the periodic branch detaches from the stationary branch. The periodic regime then organizes along isolated branches referred to as \emph{isolas} and the stationary regime remains stable across the entire range of current values considered (panels \textbf{b} and \textbf{c}). This branch of stable steady states corresponds to both resting and DB regimes, between which no clear transition occurs. In such a configuration, over its whole span of existence, the stable periodic regime coexists with the stable rest/DB stationary regime. Figure~S5 illustrates this bistability in the case $I_{\rm app} = \qty{20}{\micro\ampere\per\centi\meter\squared}$. 
Note that $s_{\rm tot}$, the weighted sum of the slow inactivation gating variables, is smaller at the stable stationary state ($s_{\rm tot}\approx0.2$) than its averaged value in the stable periodic regime ($s_{\rm tot}\approx0.42$), see panel \textbf{d}. It is also smaller than at the steady state in the absence of applied current (blue dot in Fig.~S4 (\textbf{c})). The difference is considerable, in light of the fact that $s_{\rm tot}$ evolves only very slowly. It is not clear what realistic scenario could, under natural conditions, lead the system to settle into the stable stationary regime instead of the stable periodic regime. Ramped voltage-clamp electrophysiology protocols could help determine whether such a state exists. This is an interesting question for further investigation. 

%
\subsection*{Hyperthermia}
%

Motivated by its implication in DS, notably the occurrence of febrile seizures, we investigate the impact of elevated temperature on the system's dynamics. In the model, temperature is accounted for by scaling the activation and inactivation rates of the ion channels, as is commonly done.
Hence, at higher temperatures all gating variables evolve on faster timescales.  

The consequences of accelerated kinetics of slow inactivation were addressed in previous sections. We showed that it affects the transient dynamics of the system, by precipitating the onset of slow inactivation-mediated DB. An increase in temperature from \qty{33}{\degreeCelsius} (default) to \qty{40}{\degreeCelsius} (fever) results in the slow inactivation variables $s_{\rm wt}$ and $s_{\rm mut}$ evolving approximately twice faster: $\Phi_s(40)=2.11\,\Phi_s(33)$. In such conditions, half the time is needed for slow inactivation to reach the point where the neuron can no longer spike. 

Concerning the other gating variables, it has been shown, for the Hodgkin-Huxley model, that temperature-induced acceleration of their kinetics alters the system's bifurcation structure with respect to the applied current. Specifically, I. Labouriau~\cite{labouriau1985,labouriau1989} proved, using singularity theory, the disappearance of the two Hopf bifurcation points and the creation of isolas for the periodic regime (see also the book by Golubitsky and Schaeffer~\cite{golubitsky1985}). Following this theoretical work, these isolas were first computed a few years after~\cite{hassard1989,shiau1991}.

\begin{figure}[ht!]
\centering
\includegraphics[]{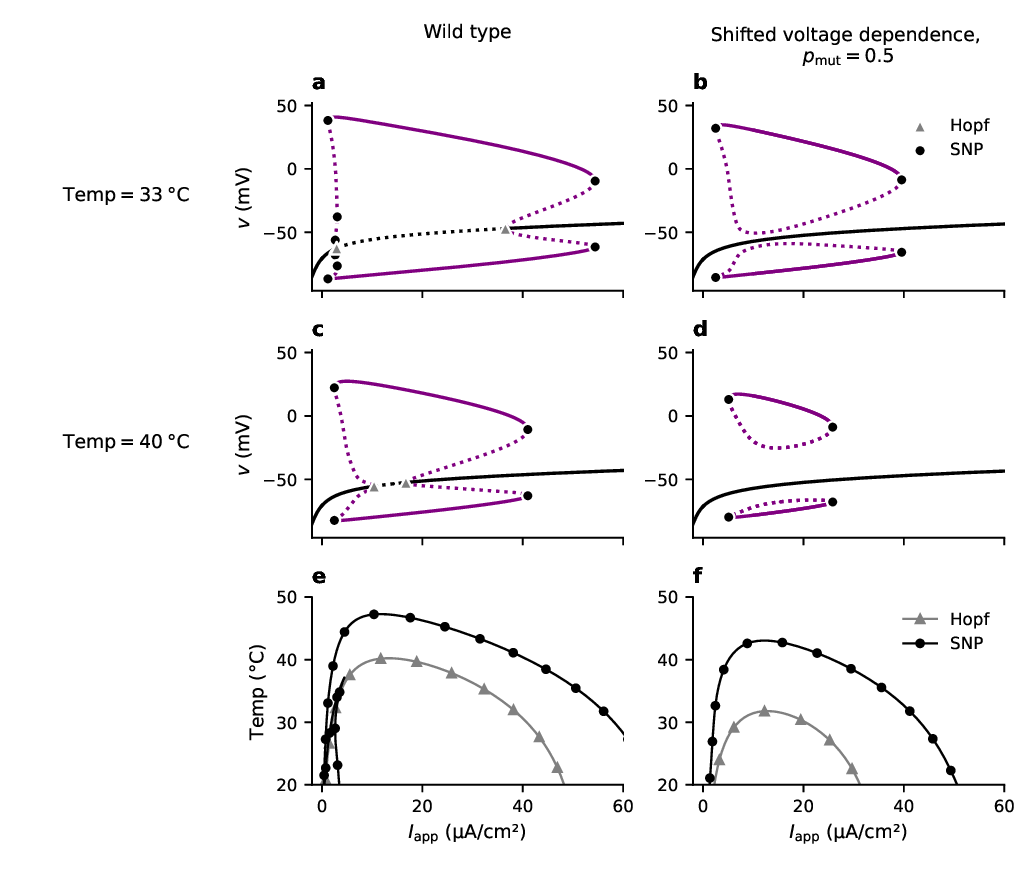}
\caption{Effect of high temperature on the asymptotic behavior of the full system. 
{\bf(a-d)} Bifurcation diagrams of the full system with respect to the applied current. 
{\bf (a)} Wild type neuron, default temperature $T=\qty{33}{\degreeCelsius}$. {\bf (b)} Shifted voltage dependence of slow inactivation ($v_s = \qty{-15}{\milli\volt}$) for half of the \na channels ($p_{\rm mut}=0.5$), default temperature. 
Note that panels {\bf a} and {\bf b} are the same as panels {\bf a} and {\bf c} of Fig.~\ref{fig:bd_wrt_iapp}, reproduced here to facilitate the comparison with the hyperthermia condition.
{\bf (c)} Wild type neuron, elevated temperature $T=\qty{40}{\degreeCelsius}$. {\bf (d)} Shifted voltage dependence of slow inactivation for half of the \na channels, elevated temperature $T=\qty{40}{\degreeCelsius}$.
{\bf(e-f)} Two-parameter bifurcation diagrams with respect to the applied current and the temperature in the two conditions: wild type (panel {\bf e}), and shifted voltage dependence of slow inactivation (panel {\bf f}).
}
\label{fig:bd_wrt_iapp_high_temp}
\end{figure}

In our Hodgkin-Huxley type model, we also obtain temperature-induced isolas of limit cycles. Figure~\ref{fig:bd_wrt_iapp_high_temp} (\textbf{a-d}) shows bifurcation diagrams of the full system with respect to the applied current, in the wild type case and when the voltage dependence of slow inactivation is shifted for half of the \na channels, at \qty{33}{\degreeCelsius} (same as in Fig.~\ref{fig:bd_wrt_iapp} (\textbf{a, c})) and at \qty{40}{\degreeCelsius} (elevated temperature).
In each case (i.e., wild type and shifted voltage dependence), we also numerically continue the Hopf and SNP points with respect to the temperature (panels \textbf{e-f}). We can see in both configurations that, 
as temperature increases, the two Hopf bifurcations approach one another, eventually colliding and giving rise to an isola. 
Meanwhile, the two SNP bifurcations stabilizing the branch of limit cycles also move closer together, reducing the range of applied current for which tonic firing is a stable solution of the full system. 
The effect of higher temperature on the long-term activity regime of the neuron is therefore similar to that of more negative $v_s$ values (Fig.~\ref{fig:bd_wrt_iapp}), exacerbating susceptibility to DB onset.

%

\section*{Discussion}

In this paper, we investigate how alterations in \nav slow inactivation, caused by a specific missense DS variant (A1783V), can impair the function of fast-spiking GABAergic neurons. We seek to better understand the initial phase of the disease, during which compromised inhibition is believed to play a central role in the pathological mechanism, prior to secondary modifications \cite{rusinaVoltagegatedSodiumChannels2023}.
Relying on simulation and bifurcation analysis of a conductance-based model incorporating experimentally measured alterations of \nav slow inactivation \cite{layerDravetVariantSCN1AA1783V2021}, we demonstrate a mechanism by which these changes may contribute to firing dysfunction.
This result contrasts with, and provides a complementary perspective, to Layer \textit{et al.}’s prediction that altered activation of the A1783V variant is the main driver of impaired firing.
Specifically, we find that the more permissive voltage-dependence and faster kinetics of slow inactivation of mutant channels lead to a more drastic reduction in channel availability upon prolonged spiking, ultimately triggering DB.  
Although we focused on a specific missense mutation, our results are also relevant in the case of truncating mutations. With such mutations, fewer functional channels are available from the outset, rather than after prolonged spiking. According to our bifurcation analysis, this suggests early increased propensity to DB.

The contribution of GABAergic neurons' DB to the onset of seizure-like activity is supported by in vitro \cite{ziburkusInterneuronPyramidalCell2006,calinDisruptingEpileptiformActivity2021a} and computational evidence \cite{kimInfluenceDepolarizationBlock2017,lemaireModelingNaV11SCN1A2021}.
Additionally, a study in a slice model of focal epilepsy suggests that DB of GABAergic neurons promotes seizure propagation \cite{cammarotaFastSpikingInterneuron2013}.
The hypothesized mechanism is as follows: when GABAergic neurons enter DB, the cessation of firing disrupts neurotransmitter release, resulting in the failure of their inhibitory restraint \cite{lignaniUnblockBlockPreventing2022}.
Notably, Călin \emph{et al.} \cite{calinDisruptingEpileptiformActivity2021a} showed that preventing DB selectively in PV+ interneurons reduces the probability of initiating epileptiform activity in vitro.
Experimental findings suggest that increased susceptibility of GABAergic neurons to DB may be involved in DEEs other than DS \cite{bereckiSCN1AGainFunction2019,mirallesParvalbuminInterneuronImpairment2024}.
Furthermore, although not always explicitly discussed, DB can be observed in several voltage traces of GABAergic neurons from DS mice, including A1783V mouse models \cite{kuoDisorderedBreathingMouse2019,diberardinoTemporalManipulationScn1a2024,mattisCorticohippocampalCircuitDysfunction2022,yuanAntisenseOligonucleotidesRestore2024,lemaireModelingNaV11SCN1A2021}.
In particular, Yuan \emph{et al.} \cite{yuanAntisenseOligonucleotidesRestore2024} showed an increased sensitivity of PV+ neurons to DB in DS mice (\emph{Scn1a\textsuperscript{tm1Kea}}), which was resolved by STK-001 injection (diseases-modifying treatment to up-regulate the wild type allele), thereby restoring GABAergic signalling to pyramidal cells.
We note a difference in the timing of DB onset between the recordings from these different studies, in which it 
occurs after hundreds of milliseconds, and simulations of our model where DB occurs after tens of seconds.
At this stage, the reason for this discrepancy is unclear. A distinct mechanism than slow inactivation may 
underlie DB in the recordings. Another possibility is that slow inactivation operates overall faster than assumed in the model. 
Ultimately, the DB mechanism predicted by our model remains to be confirmed experimentally.

A mechanism of transition to the block, driven by the accumulated slow inactivation of \na channels, is plausible and has been proposed in contexts other than Dravet syndrome. Tucker \emph{et al.} \cite{tuckerPacemakerRateDepolarization2012} reported that pharmacologically decreasing \na conductance in dopamine neurons increases their susceptibility to DB, while augmenting this conductance has the opposite effect.
In a subsequent study, Qian \emph{et al.} \cite{qianMathematicalAnalysisDepolarization2014} showed that including slow \na inactivation into a computational model of dopamine neurons allows capturing the manner in which they enter DB more faithfully than models with only fast inactivation. Note that the slow-fast dissection represented in Fig.~(3c) of their paper resembles the ones in Fig.~\ref{fig:slow_fast_all_channels} (\textbf{c}) and Fig.~\ref{fig:slow_fast_tau_all_channels} (\textbf{c, d}), with the difference that our slow inactivation gating variable evolves much more slowly.
It has also been suggested that DB may arise from the slow inactivation of \na channels in olfactory sensory neurons, interestingly as part of their normal physiological function \cite{tadresDepolarizationBlockOlfactory2022}.
In an earlier study, recordings of cortical pyramidal neurons and simulations of a Hodgkin-Huxley type model indicate that the use-dependent removal of \na channels from the available pool during high frequency discharges may lead to spike failure resembling DB \cite{fleidervishSlowInactivationNa1996}. 
In the context of epilepsy, in the aforementioned study by Călin \emph{et al.} \cite{calinDisruptingEpileptiformActivity2021a}, DB could be prevented by targeting \na inactivation with
pulsed light activation of a hyperpolarizing opsin,
resulting in a lower incidence of epileptiform activity.

In this work, we modeled the effect of temperature on channel gating by scaling the transition rates with $Q_{10}$ temperature coefficients.
The effect we observe, namely the formation of isolas of limit cycles at larger temperature, is in agreement with what is obtained in the Hodgkin-Huxley model when temperature is modeled in the same way \cite{labouriau1985,labouriau1989,golubitsky1985,hassard1989,shiau1991}.
Related to this phenomenon, our model also shows increased sensitivity to DB at higher temperatures.
However, the impact of temperature on channel gating may be more complex.
For example, temperature may affect the steady state voltage-dependence of the activation and inactivation mechanisms, in ways that can differ between wild type and mutant channels \cite{jonesL1624QVariantSCN1A2021,petersTemperaturedependentChangesNeuronal2016}.
Notably, Peters \emph{et al.} characterized the gating properties of wild type and A1273V \nav channels at different temperatures. The latter belongs to the wider DS spectrum.
They report variant-specific shifts of the activation and fast inactivation voltage dependence at larger temperature. By implementing these shifts in a computational model, they further predict reduced DB sensitivity with the mutant channels, and at higher temperature. Interestingly, this contrasts with our predictions. To our knowledge, a similar experimental characterization of the temperature dependence of channel gating is not available for A1783V. 
Further investigation is needed to elucidate the influence of temperature on the occurrence of DB, which may depend on the specific variant involved.

Future research could proceed in different directions. First, as mentioned above, it would be highly valuable to experimentally test the mechanism of transition to DB predicted in this paper. Are fast-spiking GABAergic neurons of \emph{Scn1a}\textsuperscript{+/A1783V} mice more susceptible to DB upon prolonged stimulation? If so, is slow inactivation of \na channels the underlying mechanism, and what duration and intensity of stimulation are required to induce the transition? 
Kuo \emph{et al.} \cite{kuoDisorderedBreathingMouse2019} reported a more prone entry into DB for inhibitory neurons heterologously expressing the A1783V variant.
It would be interesting to determine whether these blocks, which occur earlier than in our model, are mediated by slow inactivation or by another mechanism.
More generally, to which extent is slow inactivation-mediated DB involved in pathological and physiological neuronal processes? 
From a modeling perspective, one could investigate whether the alteration of \nav activation, not considered in this study, contributes to DB onset. The model could also be used to test whether factors, such as the upregulation of potassium channels, could counteract the transition to DB. 
Finally, in the spirit of the work by Kim \emph{et al.} \cite{kimInfluenceDepolarizationBlock2017}, a natural next step could be to explore, in a network model, the failure of inhibitory restraint resulting from DB of GABAergic neurons, and its repercussions on the activity of glutamatergic neurons, potentially facilitating seizure initiation.
To scale up from the single neuron to the network level, it would be useful to derive a minimal neuron model of type integrate-and-fire capable of capturing the transition to DB via a suitable dynamical scenario.

\vspace{0.8cm}
\paragraph*{\textbf{Data availability.--}}
The code used to produce the figures in this manuscript is available in a GitHub repository at \url{https://github.com/louisianelemaire/Slow-NaV-inactivation-in-Dravet-syndrome} and \url{https://gitlab.inria.fr/llemaire/Slow-NaV-inactivation-in-Dravet-syndrome}.
\vspace{0.5cm}
\paragraph*{\textbf{Acknowledgments.--}}
LL, MD and FC acknowledge the support of Inria through the Exploratory Action AEX-2MDS. SR acknowledges the grant
PID2023-146683OB-100 funded by MICIU/AEI /10.13039/501100011033 and by ERDF, EU. Additionally, SR acknowledges
support from Ikerbasque Foundation and the Basque Government through the BERC 2022-2025 program and by the Ministry of Science and Innovation: BCAM Severo Ochoa accreditation CEX2021-001142-S / MICIU / AEI / 10.13039/501100011033.
Moreover, SR acknowledges the financial support received from BCAM-IKUR, funded by the Basque Government by the
IKUR Strategy and by the European Union NextGenerationEU/PRTR, as well as, support of ONBODY no. KK-2023/00070
funded by the Basque Government through ELKARTEK Programme.

\bibliography{dravet}

\newpage

\onecolumngrid
\renewcommand{\theequation}{S\arabic{equation}} 
\renewcommand{\thefigure}{S\arabic{figure}}
\renewcommand{\thesection}{S\arabic{section}}
\setcounter{equation}{0}  
\setcounter{section}{0}  
\setcounter{figure}{0}  

\begin{center}
	\fontsize{20}{24}\bfseries\scshape{Supplementary Information (SI)}
\end{center}

\section*{Gating functions of the model}
The gating functions associated with the model we are studying (System (1) of the main text) are given by:
\begin{nalign}
		\alpha_m & = 0.2567 \, \frac{-(v-v_v + 60.84)}{\exp{\left(\frac{-(v-v_v + 60.84)}{9.722}\right)}-1}\,,
		&
		\beta_m & = 0.1133 \,\frac{v-v_v + 30.253}{\exp{\left(\frac{v-v_v + 30.253}{2.848}\right)}-1}\,,
	\\
		\alpha_h & = 0.00105\, \exp{\left(\frac{-(v-v_v)}{20}\right)}\,,
		&
		\beta_h &= 4.827 \,\frac{1}{\exp{\left(\frac{-(v-v_v+18.646)}{12.452}\right)} + 1}\,,
	\\
		\alpha_n & = 0.0610 \,\frac{-(v-29.991)}{\exp{\left(\frac{-(v-29.991)}{27.502}\right)}-1}\,,
		&
		\beta_n &= 0.001504 \,\exp{\left(\frac{-v}{17.177}\right)}\,,
	\\
		\alpha_{\tilde{n}} & = 0.0993 \,\frac{-(v-33.720)}{\exp{\left(\frac{-(v-33.720)}{12.742}\right)}-1}\,,
		&
		\beta_{\tilde{n}} &= 0.1379\, \exp{\left(\frac{-v}{500}\right)}\,,
	\\
		x_{\infty} & = \frac{\alpha_x}{\alpha_x + \beta_x},\;x\in\{m,h,n,\tilde{n}\}\,,
		&
		\tau_{x} &= \frac{1}{\alpha_x + \beta_x},\;x\in\{h,n,\tilde{n}\}\,,
	&
        s_{\infty} & = \frac{1}{1+\exp{\left(\frac{-(v - v_h)}{k}\right)}}\,.
\end{nalign}

\section*{Parameter values of the model}
The parameter values we have used to simulate the model and compute various bifurcation diagrams displayed in the main text, are gathered in the table below.
\renewcommand{\arraystretch}{1.2}
\begin{table}[ht]
\centering
\begin{tabular}{|l|l|l|l|l|}
\hline
Symbol & Description & Value & Unit & Source \\
\hline\hline
$I_{\rm app}$ & Applied current & & \unit{\micro\ampere\per\centi\meter\squared}   & \\
\hline
C & Membrane capacitance & 0.9 & \unit{\micro\farad\per\centi\meter\squared} & \cite{huComplementaryTuningNa2018} \\
\hline
$g_{\rm Na}$ & Na\textsuperscript{+} maximal conductance & 70 & \unit{\milli\siemens\per\centi\meter\squared} & \\
\hline
$g_{\rm K}$ & K\textsuperscript{+} maximal conductance & 15 & \unit{\milli\siemens\per\centi\meter\squared} & \cite{huComplementaryTuningNa2018}\\
\hline
$g_{\rm leak}$ & Leak conductance & 0.1 & \unit{\milli\siemens\per\centi\meter\squared} & \cite{huComplementaryTuningNa2018}\\
\hline
$v_v$ & Correction of the voltage dependence offset after patch excision & 20 & \unit{\milli\volt} & \cite{huComplementaryTuningNa2018}\\
\hline
$v_s$ & Na\textsuperscript{+} slow inactivation voltage dependence offset & 0, -15 & \unit{\milli\volt} & \cite{layerDravetVariantSCN1AA1783V2021} \\
\hline
$v_h$ & Steady state Na\textsuperscript{+} slow inactivation midpoint voltage & -60 & \unit{\milli\volt} & \cite{layerDravetVariantSCN1AA1783V2021} \\
\hline
$k$ & Steady state Na\textsuperscript{+} slow inactivation slope parameter & -10 & \unit{\milli\volt} & \cite{layerDravetVariantSCN1AA1783V2021} \\
\hline
$\tau_{s}$ & Na\textsuperscript{+} slow inactivation time constant & 30000, 3000 & \unit{\milli\second} & \cite{layerDravetVariantSCN1AA1783V2021} \\
\hline
$E_{\rm Na}$ & Na\textsuperscript{+} reversal potential & 55 & \unit{\milli\volt} & \cite{huComplementaryTuningNa2018}\\
\hline
$E_{\rm K}$ & K\textsuperscript{+} reversal potential & -90 & \unit{\milli\volt} & \cite{huComplementaryTuningNa2018}\\
\hline
$E_{\rm leak}$ & Leak reversal potential & -65 & \unit{\milli\volt} & \cite{huComplementaryTuningNa2018}\\
\hline
$Q_{10, m}$ & Temperature coefficient for Na\textsuperscript{+} activation & 2.2 & - & \cite{huComplementaryTuningNa2018} \\
\hline
$Q_{10, h}$ & Temperature coefficient for Na\textsuperscript{+} fast inactivation & 2.9 & - & \cite{huComplementaryTuningNa2018} \\
\hline
$Q_{10, n}$ & Temperature coefficient for K\textsuperscript{+} activation & 3 & - & \cite{huComplementaryTuningNa2018} \\
\hline
$Q_{10, s}$ & Temperature coefficient for Na\textsuperscript{+} slow inactivation & 2.9 & - &  \\
\hline
$T$ & Temperature & 33 & \unit{\degreeCelsius} & \cite{layerDravetVariantSCN1AA1783V2021} \\
\hline
\end{tabular}
\caption{Model parameters.}
\label{tab:model_params}
\end{table}
\newpage
\section*{Supplementary figures}
\begin{figure}[ht!]
\centering
\includegraphics[]{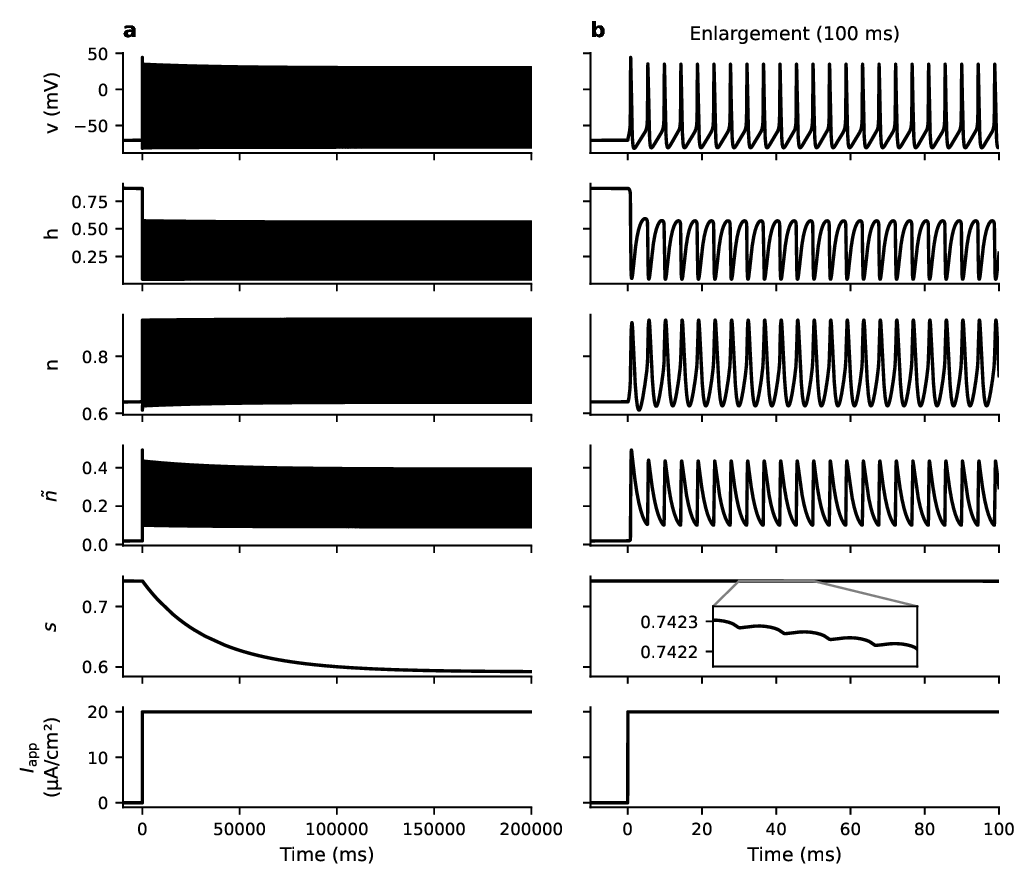}
\caption{
\textbf{(a)} Time traces of all state variables of System (1) of the main text, in response to a $\qty{200}{\second}$ current step of $\qty{20}{\micro\ampere\per\centi\meter\squared}$. Wild type configuration: $v_s=\qty{0}{\milli\volt}$, $\tau_s=\qty{30000}{\milli\second}$. It corresponds to the green trajectory in the fourth panel of Fig. 1 of the main article.
\textbf{(b)} First $\qty{100}{\milli\second}$ of simulation. Note that $s$ is almost constant.
}
\end{figure}

\begin{figure}[ht!]
\centering
\includegraphics[]{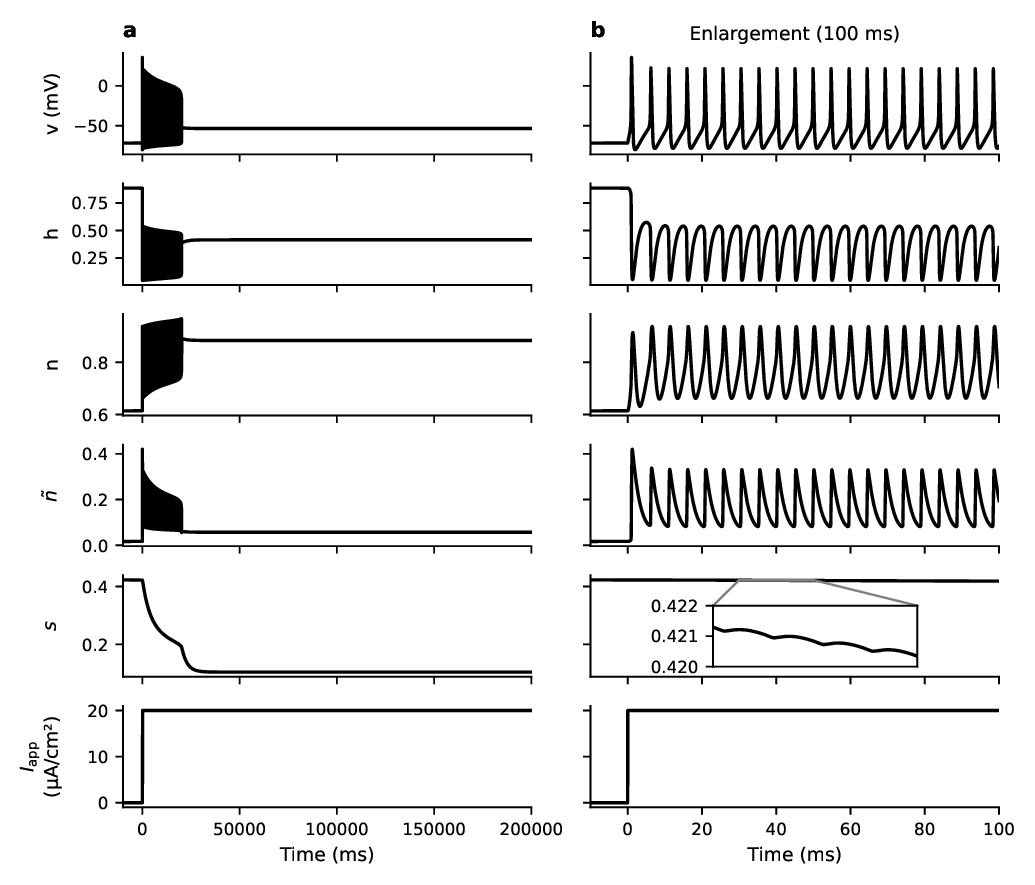}
\caption{
\textbf{(a)} Time traces of all state variables of the model defined in (1), in response to a $\qty{200}{\second}$ current step of $\qty{20}{\micro\ampere\per\centi\meter\squared}$. Configuration with enhanced slow inactivation for all sodium channels: $v_s=\qty{-15}{\milli\volt}$ (voltage dependence), $\tau_s=\qty{3000}{\milli\second}$ (kinetics). It corresponds to the red trajectory in the fourth panel of Fig. 1 of the main article.
\textbf{(b)} First $\qty{100}{\milli\second}$ of simulation. Note that $s$ is almost constant.
}
\end{figure}
\begin{figure}[ht!]
\centering
\includegraphics[]{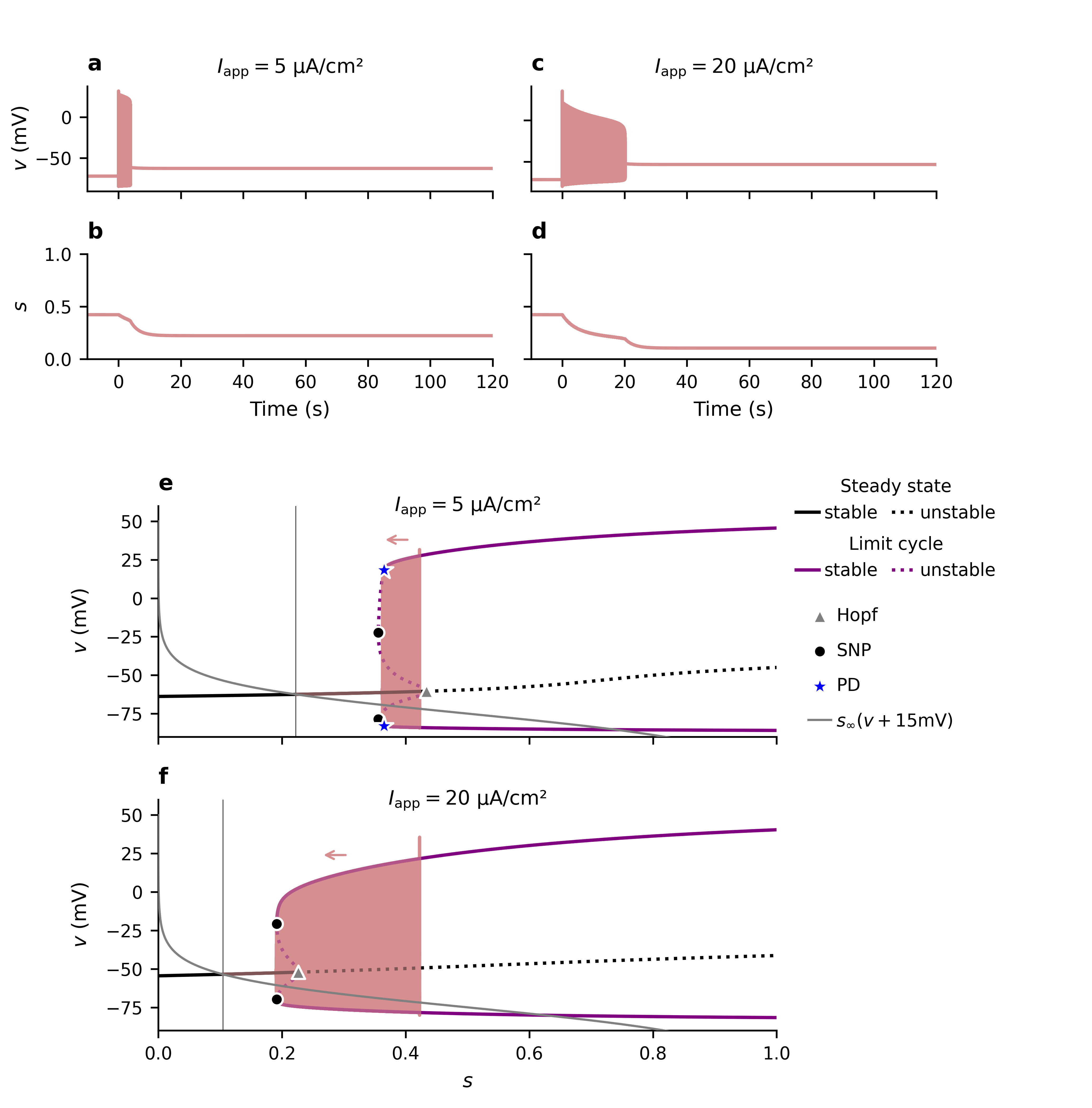}
\caption{
Resting state when $I_{\rm app} = \qty{0}{\micro\ampere\per\centi\meter\squared}$.
Voltage {\bf (a)} and $s$ {\bf (b)} steady states when no current is applied {\bf (b, right axis)}, in the wild type case (${v}_s = \qty{0}{\milli\volt}$; green curves) or with the shift of the voltage dependence of slow inactivation (${v}_s = \qty{-15}{\milli\volt}$; blue curves).
{\bf (c)} Bifurcation diagram of the fast subsystem with respect to $s$ when $I_{\rm app} = \qty{20}{\micro\ampere\per\centi\meter\squared}$.
{\bf (d)} Time derivative of the slow variable $s$ on stable steady states of the fast subsystem, without (upper curve) or with (lower curve) the shift of $s_{\infty}$.
}
\end{figure}
\begin{figure}[ht!]
\centering
\includegraphics[]{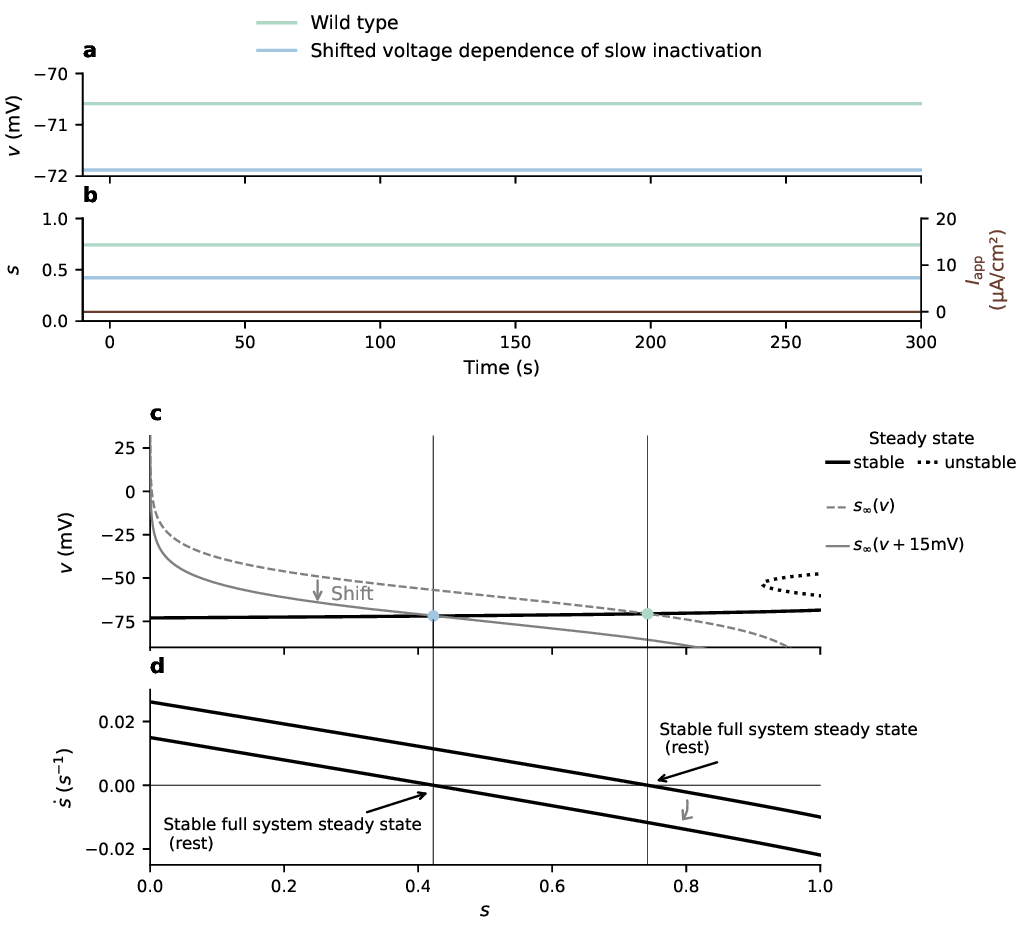}
\caption{
Bistability of the full system between a periodic and a stationary regime; case study when $I_{\rm app} = \qty{20}{\micro\ampere\per\centi\meter\squared}$ and $p_{\rm mut}=0.5$. Slow inactivation alteration: shifted voltage dependence with ${v}_s = \qty{-15}{\milli\volt}$, default kinetics with $\tau_s=\qty{30000}{\milli\second}$. 
Setting $\tau_{s, {\rm mut}}=\tau_{s, {\rm wt}}$ does not influence which limit activity regimes exist for the neuron, yet it simplifies the slow–fast dissection.
It allows us to reduce slow dynamics to one dimension, by considering as slow variable $s_{\rm tot}$ instead of $s_{\rm wt}$ and $s_{\rm mut}$. The averaged slow subsystem is given by
    $\dot{\langle s_{\rm tot}\rangle} = \frac{\Phi_s({\rm Temp})}{\tau_{s, {\rm wt}}} \left( \int_0^{T(s_{\rm tot})} p_{\rm wt}\, s_\infty\bigl(v(s_{\rm tot}, \tau)\bigr) + p_{\rm mut}\, s_\infty\bigl(v(s_{\rm tot}, \tau) - v_s\bigr) \,{\rm d}\tau -s_{\rm tot} \right)$.
    {\bf (a-b)} Voltage and $s$ time traces in response to a two-minute current step {\bf (b, right axis)}. {\bf (a, right axis)} Instantaneous firing frequency (darker colored lines). {\bf (c)} Full system trajectories projected onto the bifurcation diagram of the fast subsystem with respect to $s_{\rm tot}$ when $I_{\rm app} = \qty{20}{\micro\ampere\per\centi\meter\squared}$.
    {\bf (d)} Time derivative of the slow variable $s_{\rm tot}$ on stable steady states of the fast subsystem or averaged time derivative of $s_{\rm tot}$ on stable limit cycles of the fast subsystem. 
}
\end{figure}

\begin{figure}[ht!]
\centering
\includegraphics[]{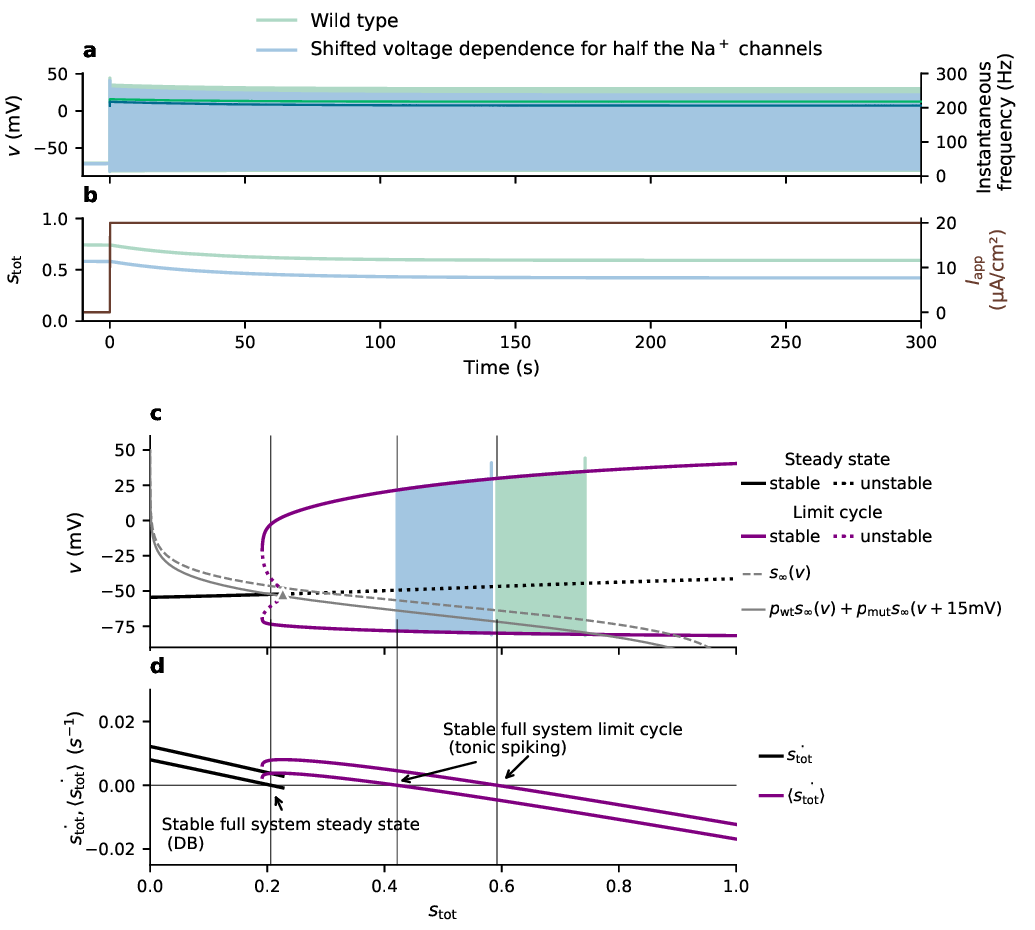}
\caption{
Bistability of the full system between a periodic and a stationary regime; case study when $I_{\rm app} = \qty{20}{\micro\ampere\per\centi\meter\squared}$ and $p_{\rm mut}=0.5$. Slow inactivation alteration: shifted voltage dependence with ${v}_s = \qty{-15}{\milli\volt}$, default kinetics with $\tau_s=\qty{30000}{\milli\second}$. 
Setting $\tau_{s, {\rm mut}}=\tau_{s, {\rm wt}}$ does not influence which limit activity regimes exist for the neuron, yet it simplifies the slow–fast dissection.
It allows us to reduce slow dynamics to one dimension, by considering as slow variable $s_{\rm tot}$ instead of $s_{\rm wt}$ and $s_{\rm mut}$. The averaged slow subsystem is given by
    $\dot{\langle s_{\rm tot}\rangle} = \frac{\Phi_s({\rm Temp})}{\tau_{s, {\rm wt}}} \left( \int_0^{T(s_{\rm tot})} p_{\rm wt}\, s_\infty\bigl(v(s_{\rm tot}, \tau)\bigr) + p_{\rm mut}\, s_\infty\bigl(v(s_{\rm tot}, \tau) - v_s\bigr) \,{\rm d}\tau -s_{\rm tot} \right)$.
    {\bf (a-b)} Voltage and $s$ time traces in response to a two-minute current step {\bf (b, right axis)}. {\bf (a, right axis)} Instantaneous firing frequency (darker colored lines). {\bf (c)} Full system trajectories projected onto the bifurcation diagram of the fast subsystem with respect to $s_{\rm tot}$ when $I_{\rm app} = \qty{20}{\micro\ampere\per\centi\meter\squared}$.
    {\bf (d)} Time derivative of the slow variable $s_{\rm tot}$ on stable steady states of the fast subsystem or averaged time derivative of $s_{\rm tot}$ on stable limit cycles of the fast subsystem. 
}
\end{figure}

\end{document}